# Robust Sandwiched B/TM/B Structures by Metal Intercalating into Bilayer Borophene Leading to Excellent Hydrogen Evolution Reaction

Yuan Chang[1,2], Jiaxu Liu[3*], Hongsheng Liu[1,2], Yong-Wei Zhang[4], Junfeng Gao[1,2*], Jijun Zhao[1,2*]

1. *State Key Laboratory of Structural Analysis for Industrial Equipment, Dalian University of Technology, Dalian, 116024, China*
2. *Laboratory of Materials Modification by Laser, Ion and Electron Beams, Ministry of Education, Dalian University of Technology, Dalian, 116024, China*
3. *State Key Laboratory of Fine Chemicals, School of Chemical Engineering, Dalian University of Technology, Dalian 116024, P. R. China.*
4. *Institute of High Performance Computing (IHPC), A\*STAR, Singapore, 138632, Singapore*

\*Email: liujiaxu@dlut.edu.cn; gaojf@dlut.edu.cn; zhaojj@dlut.edu.cn

**Abstract.** Bilayer borophene, very recently synthesized on Ag and Cu, possesses extremely flat large surface and excellent conductivity. The van der Waals gap of bilayer borophene can be intercalated by metal atoms, tailoring the properties of bilayer borophene. Herein, we propose sandwiched B/TM/B (TM=Co, Ni, Cu, Pd) is a new 2D formation by transiton metal atoms intercalated into bilayer borophene, it is quite robust with both energetic, structural and thermal stability, and exhibits heat resistance of at least 1300 K. Besides, it is novel platform for electrocatalytic hydrogen evolution reaction (HER). The interecalation metal atom serves as single-atomic catalyst. Beyond, the transtion metal is protected by outside boron layers from being corroded by acidic/alkaline solution. $B/Cu_x/B$, $B/Pd_x/B$ and $B/Al_x/B$ with different metal coverage exhibit defect-independent extremely low HER free energy in the range of -0.162 ~ 0.179 eV, -0.134 ~ 0.183 eV and -0.082 ~ 0.086 eV which are comparable to noble metal Pt. Combining excellent conduction, high structural and thermal stability, low resistance to intercalated behaviour, effortless water splitting process, excellent defect-independent catalytic performance, cheapness and abundance of raw materials, free of corrodation, 2D sandwiched B/TM/B (TM=Co, Ni, Cu, Pd) is believed to promising for electrocatalytic HER applications.

# 1. Introduction

After the industrial revolution, the massive burning of fossil fuels has led to worrisome environmental and climatic concerns, calling in clean and renewable energy sources. Hydrogen evolution reaction (HER), a branch of water splitting reaction, is believed to present a possible solution.[1] Noble metal Pt catalyst is superior for electrocatalytic HER but limited by its high expense.[2] Two-dimensional (2D) materials with large surface are expected to be substituted for Pt as next generation of catalytic platform. Several 2D materials without Pt, such as $MoS_2$,[3-11] $WS_2$,[12,13] $Mo_2C$,[14-16] MoP,[17,18] FeP,[19] NiMo alloy[20] and MOF[21] have been explored, opening up a feasible way for HER.[22] However, the active sites in those catalysts mainly locate at defects and edges, thus limiting their practical applications. In addition, the conductivity of electrode material plays a decisive role in the electrocatalytic performance. Thus, an ideal candidate material must meet the following essential requirements to reach excellent electrocatalytic performance: (1) large and flat surface, (2) multiple sites on the surface with high chemical activity independent of defects, (3) good electrical conductivity, (4) robust against to corrodation.

Borophene is considered as the new wonder material after graphene,[23,24] exhibiting superior electrical conductivity better than graphene,[25,26] lattice thermal conductivity,[27-29] inertness to oxidation[30] and large anisotropic tensile strength.[31-34] Therefore, borophene is ideal modifiable material meeting the structural and conductive requirements for good electrocatalysis. Monolayer borophene has been achieved on several metal substrates.[35-45] Very recently, by breaking the self-limiting effect, metallic bilayer borophene has been successfully fabricated on Ag(111)[46] and Cu(111) surface.[47] However, both monolayer and bilayer borophene must be supported on metal substrate due to their intrinsic electron deficiency.

Except for charge transferring from metal substrate, regulation like metal doping and intercalation can be also pervaded ways to afford external charge required for the stability of borophene. Several freestanding metal-borophene networks including the second and third main group elements (Mg,[48] Ca,[48] Sr,[49] Al,[50-53] Ga,[54] In[54]) as well as partial transition metal elements (Sc,[48,55] Ti,[48,56-58] V,[56] Cr,[56] Mn,[56,59,60] Fe,[61,62] Y,[48] Zr,[57,58] Mo,[63] Hf,[57,58] W,[64] Ir[65]) have been reported. Among the transition metal borides (TMB) coexist various attractive characteristics, especially, strong mechanical properties and stability. For instance, CrB-type MnB combines high mechanical hardness and ferromagnetism,[59] another tpye TMB $XB_2$ (X=Ti, Zr, Hf) break through the challenge of coexistence of conductivity, strong mechanical properties, and hydrophobicity in one material.[48] In catalytic applications, these freestanding networks burst out stronger flexibility and portability than those substrate-based ones. On the other hand, the metal doped or intercalated systems may exhibit special single atom catalysis, where the active metal atoms are wrapped and protected, but the electrons from metal atoms can penetrate the covering layer to play a catalytic role. For example, appreciable catalytic performance can be obtained by intercalation of sandwich-like X/M/X structure of transition metal dichalcogenide,[10-13] layered double hydroxides[66-68] and Mxenes.[16,69-71]

Intercalation has been experimentally proved to be feasible in common 2D materials including graphene,[72] phosphorene[73] and MoS$_2$,[74,75] but still needs further study for commercialization. Borophene is excellent conductor, Fermi velocity of carriers is even faster than graphene with Dirac Fermions. Intercalation bilayer borophene may transform it into an ideal electrochemical catalyst. Therefore, one might raise the question that are there other 2D metal-borophene networks except for those reported previously?[48-65] What is the most stable structure? Especially, the potential catalytic application of metal-borophene networks is an interesing issue, which may offer more opportunities to search next generation of catalytic platform.

Herein, highly stable transition metal intercalated bilayer borophene networks, B/TM/B (TM=Co, Ni, Cu, Pd), are proposed by high throughput calculations. We emphasize twenty transition metals in the third and fourth period elements due to their complex arrangement of d-orbital electrons and aplenty exotic properties. Our models comes from global structure search by CALYPSO code,[76,77] which possess the lowest ground state energy and manifest the same sandwiched configuration as B/M/B (M=Al, Ga, In),[52-54] which can be regards as new formation of 2D materials. The various metal coverage, water splitting process, hydrogen adsorption, charge distribution, thermal stability and kinetic behaviour of B/TM/B (TM=Co, Ni, Cu, Pd) as well as B/Al/B have been systematically investigated. B/TM/B (TM=Co, Ni, Cu, Pd) is ideal platform for water splitting with ultralow barriers. Particularly, B/Cu$_x$/B, B/Pd$_x$/B and B/Al$_x$/B with different metal coverage possess defect-independent extremely low HER free energy in the range of -0.162 ~ 0.179 eV, -0.134 ~ 0.183 eV and -0.082 ~ 0.086 eV, respectively. In B/TM/B, metal electronic states can penetrate borophene to achieve attractive catalytic performance of HER while intercalated metals are protected by wrapped borophene. Thus, the 2D sandwiched B/TM/B structure is very robust away to corrosion. Besides, B/TM/B is a high temperature resistant material, for examples, B/Ni/B and B/Pd/B exhibit the highest and lowest decomposition temperature ($T_d$) of 1750 K and 1300 K, respectively. For metal intercalation, the migration of metal from surface to interlayer of borophene bilayer is absolutely dominant process and is supposed to be the most likely formation mechanism. Interaction metal atom plays single-atom catalyst center while non-metallic borophene plays an ideal coating role with both chemical reaction inertness and God-given high conductivity. Hence, the robust sandwiched B/TM/B structures are competitive candidates for electrocatalytic platform.

## 2. Results and Discussion

### 2.1. 2D sandwiched B/TM/B formations

The intrinsic electron deficiency of borophene leads to the formation of three-center two-electron bonds (3c-2e, Figure 1a), which generally requires the introduction of external charges to maintain stable freestanding structures. There is charge transfer between metal substrates and boron atoms, for instance, Al(111) surface is beneficial for the growth of graphene-like honeycomb borophene monolayer due to abundant charge tranfer.[44] Very recently, borophene bilayer (BL) has been successfully synthesized on Cu(111),[47] where Cu doping can improve the charge distribution of borophene as well (Figure 1b). With the assistance of high throughput computation and CALYPSO code,[76,77] a new sandwiched B/Cu/B structure is determined to have exotic electronic properties and lattice stability (Figure 1c,d).

B/Cu/B exhibits a trilayer structure composed of two δ$_4$ borophene layers and one metal interlayer. The central hollow site of two boron hexagons belong to upper and lower sheets is the most stable intercalation site for metal atom. By scanning potential energy surface, binding energy of B/Cu/B on Cu(111) surface within -0.178 ~ -0.176 eV/Å$^2$ is independent of their relative position (More details in supporting information). This value of binding energy is extremely similar to -0.143 eV/Å$^2$ of Cu(111)/borophene BL.[47] Most importantly, freestanding B/Cu/B network still remains structural stability while borophene BL suffering deformation (Figure 1e,f).

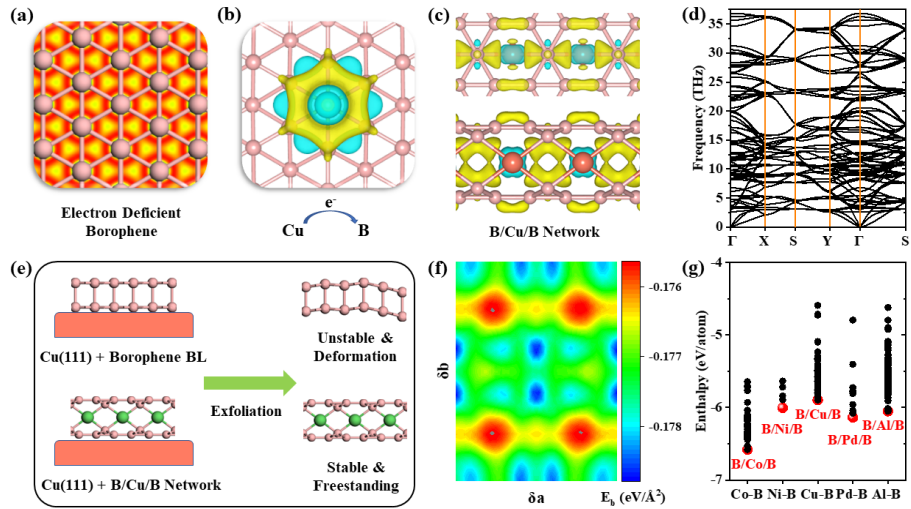

**Figure 1.** (a) Intrinsic electron deficiency of borophene. (b) Charge transfer from metal to boron and (c) charge distribution of B/Cu/B network (isosurface level = 0.006 |e|/bohr$^3$). Yellow and blue indicate electron accumulation and depletion. (d) Phonon dispersion of B/Cu/B. (e) Supported and freestanding diagrams of borophene bilayer and B/Cu/B. (f) Binding energy of B/Cu/B on Cu(111) surface (Unit: eV/Å$^2$). $E_b=(E_{total}-E_{sub}-E_{B/Cu/B})/S$. (g) Enthalpy of probable metal-borophene networks searched by CALYPSO.

Due to the third and fourth period transition elements possess complex arrangement of d-orbital electrons and aplenty exotic properties, high throughput computation is performed to search new formation. Results of B/TM/B (TM=Co, Ni, Cu, Pd) reveals their structural stability while the rest sixteen structures B/TM/B (TM=Sc, Ti, V, Cr, Mn, Fe, Zn, Y, Zr, Nb, Mo, Tc, Ru, Rh, Ag, Cd) are considered to be unstable due to the appreance of imaginary frequency in their phonon dispersion. B/Al/B [52,53] exhibits the same structural characteristic as well, thus is calculated for comparison. Besides, structure search indicates the sandwiched B/M/B network exhibit the lowest ground state energy (M=Co, Ni, Cu, Pd, Al, Figure 1g), which attracted our interest. Details are given in Figure S1, S2 and Table S1 (Supporting Information).

During the process of introducing foreign electrons using metal intercalation, borophene BL may take different numbers of intercalated atoms. As for B/TM$_x$/B (x denotes metal coverage), the formation energy is defined as

$$E_{form} = (E_{B/TM_x/B} - N_B E_B - N_M E_M)/N_M \qquad (1)$$

where $E_{B/TM_x/B}$ represents the total energy of metal-intercalated δ$_4$ borophene. E$_B$ denotes the energy of single B atom in δ$_4$ borophene while E$_M$ denotes the energy of every metal atom in its most stable bulk phase (Table S2, Supporting Information). Mutiple structures under each coverage are constructed and carefully optimized. For instance, 134 structures are constructed to determine the evolution of B/Cu$_x$/B with metal coverage x. Also, there are 112, 77, 97 and 77 structures tested for B/Co$_x$/B, B/Ni$_x$/B, B/Pd$_x$/B and B/Al$_x$/B, respectively. Detailed structures and energies are available in Figure S3~7 and Table S3~7 (Supporting Information).

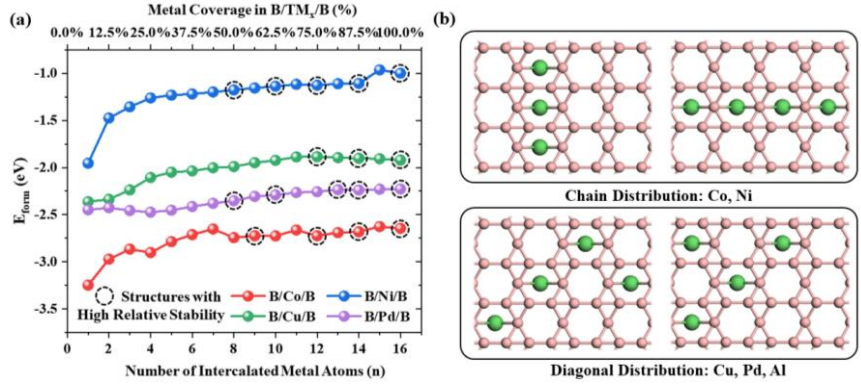

**Figure 2.** (a) Formation energy of B/TM/B (Unit: eV). Points with black dashed circle represent structures with high relative stability. (b) Typical intercalated metal distribution pattern in bilayer borophene.

According to our results, B/TM$_x$/B always have negative formation energy (Figure 2a), indicating their energetic stability. With increasing coverage of intercalated atoms, E$_{form}$ generally keeps a rising trend. The occasional abnormal drops can be attributed to the undulation and glide of δ$_4$ borophene sheets under minor intercalation. Red points in Figure S3~7 (Supporting Information) denote the most stable structures under the given coverage, from which the evolution of intercated metal atoms in B/TM$_x$/B can be determined (Figure 2b). For instance, Co atoms preferentially tend to form a framework like B/Co$_x$/B (x=75%), and subsequently inserted atoms patch the pores of the framework. Ni atoms obey linear distribution in bilayer borophene. Pd atoms will form a grid along diagonal direction at lower coverage, then tend to gather together at higher coverage. Al atoms prefer to distribute along diagonal direction like Pd. Cu atoms exhibit no obvious rule at lower coverage but gather together as well with increased intercalation concentration. Besides, Cu atoms lead to the most evident surface undulation among the four metal borophene networks.

Next, the stability of B/TM$_x$/B is estimated by calculating the second-order energy difference $\Delta^2 E$, which is defined as

$$\Delta^2 E = E_{N+1} + E_{N-1} - 2E_N \tag{2}$$

where N represents the number of intercalated metal atoms. Structures correspond to those positive peaks possess relatively higher stability according to our definition. The undulation and glide of borophene under minor intercalation leads to the irregular energy difference of B/TM$_x$/B, therefore have less reference value. After half coverage, nearly no futher geometry distortion occurs. For instance, B/Cu$_x$/B (x=6.25%, 18.75%, 75%, 87.5%) all exhibit relatively higher stability, while interlayer space and lattice of B/Cu$_x$/B (x=75%, 87.5%) are more uniform and regular (Figure S5, Supporting Information). Besides, several B/TM$_x$/B structures such as B/Co$_x$/B (x=56.25%, 75%, 87.5%), B/Ni$_x$/B (x=50%, 62.5%, 75%, 87.5%), and B/Pd$_x$/B (x=50%, 62.5%, 81.25%, 87.5%) all exhibit higher relative stability due to their peculiar distribution of intercalated metal atoms. Above structures is highlighted by black dashed circle in Figure 2a. In general, intercalated metal obeys chain distribution and diagonal distribution with increasing metal coverage. The practical growth pattern and kinetic behaviour can be further explained by this micro mechanism. Our results show that it is the ionic bond between the metal and B atoms that enhances the interaction between the two boron layers, thereby stabilizing the freestanding B/TM/B.

## 2.2 Water dissociation and catalytic performance

Metal adsorption could enhance the catalytic activity of borophene.[78] Considering the advantages of the large surface and numerous adsorption sites in B/TM/B, the catalytic characteristics are carefully examined. The acidity and alkalinity of an electrolyte play improtant roles in the performance of a catalyst during electrochemical processes due to the complex interactions between water molecules and metal surface.[79] Under acidic environment (usually HClO$_4$ or H$_2$SO$_4$ electrolyte), there are H$^+$ in the aqueous solution, and the process of HER can be simply described as:

$$* + H^+ + e^- \rightarrow {}^*H \tag{3}$$

$$^*H + H^+ + e^- \rightarrow H_2 + * \tag{4}$$

The process of HER under the alkaline environment (usually KOH electrolyte) can be described by the following relations:

$$H_2O + * + e^- \rightarrow {}^*H + OH^- \tag{5}$$

$$H_2O + {}^*H + e^- \rightarrow H_2 + OH^- + * \tag{6}$$

According to above relations, there are readily available protons in acidic environment. Alkaline environment requires water as proton source and need additional water molecule dissociation process, which will significantly decrease the reaction rate of the HER. Even for excellent Pt catalyst, its reaction rate is usually 2~3 orders lower in an alkali than that in an acid.[80] Therefore, efficient HER under alkaline environment requires active sites of catalyst simultaneously accelerate the water dissociation and hydrogen combination reactions.[20]

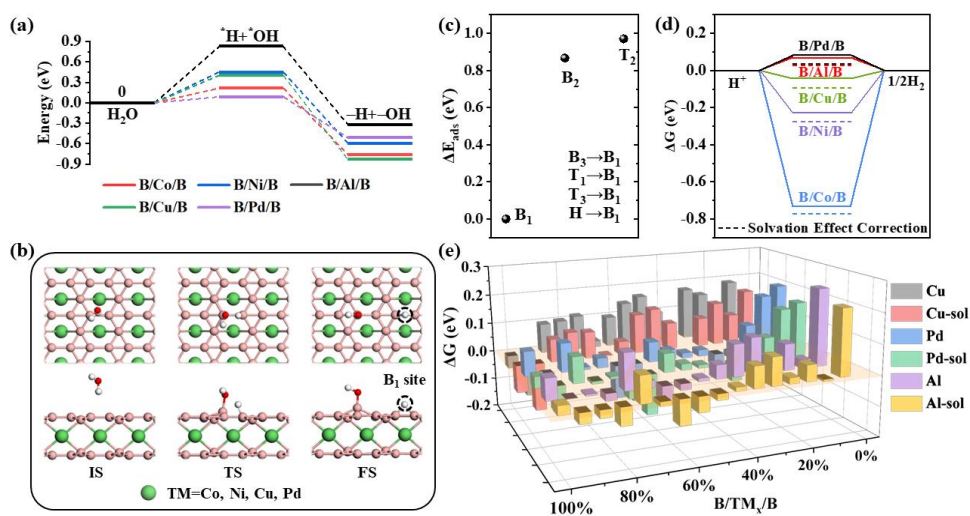

**Figure 3.** (a) Barrier of water molecule dissociation on the surface of B/TM/B (Unit: eV). (b) Diagram of dissociation process. IS, TS and FS denote the initial state, transition state and final state, respectively. (c) Hydrogen adsorption site in B/TM/B. (d) HER performance of full covered B/TM/B (Unit: eV). The dashed lines are results within water solvation effect. (e) HER performance of B/$Cu_x$/B, B/$Pd_x$/B and B/$Al_x$/B under different metal coverage (Unit: eV).

First, the process of water dissociation is investiagted. On the surface of traditional Pt catalyst, this barrier ranges from 0.44 eV of Pt(110) to 0.78 eV of Pt(111).[81-83] According to our calculations (Figure 3ab), the dissociation barrier of water molecule on B/Al/B is about 0.825 eV, which is comparable to Pt(111). B/Ni/B and B/Cu/B exhibit barriers of 0.425 eV and 0.424 eV, respectively, which are comparable to Pt(110). Meanwhile, B/Co/B shows barrier of 0.213 eV, only half of Pt(110) value. B/Pd/B exhibits an ultralow barrier of 0.087 eV for water dissociation. Detailed results are given in Figure S8 and Table S8 (Supporting Information). Consequently, B/TM/B is good platform for water dissociation and is expected to have similar or superior catalytic performance to noble Pt. Besides, *H may form hydrogen bonds with water molecules of aqueous solution, which results in deviation of theoretical prediction from experiments. Thus, the water solvation effect is considered for correction in our calculations.

Except for electrolyte and water solvation effect, another major factor is also closely related to the catalytic performance: the most stable adsorption site for H atom. Seven probable sites including bridge, top and hollow sites are tested for H adsorption. Only three sites are found to be stable after optimization (Figure 3c). Detailed results are given in Figure S9 (Supporting Information). Site $B_1$, that is, the bridge site of B hexagons, exhibits an adsorption energy about 1 eV lower than the other two stable sites, and hence it is the most stable adsorption site. Here, the adsorption energy is defined as

$$E_{ads(H*)} = E_{H*} - E_* - \frac{1}{2}E_{H_2} \qquad (7)$$

where $E_{H*}$, $E_*$ and $E_{H_2}$ are the energy of hydrogen adsorbed B/TM$_x$/B, B/TM$_x$/B and H$_2$, respectively. As for $B_1$ site, once a hydrogen atom is adsorbed there, two B-H bonds with the same length of 1.36 Å will form, and the plane where they are located is perpendicular to the borophene sheets.

The free energy ΔG are employed to evaluate the catalytic performance of B/TM$_x$/B, which is defined as

$$\Delta G = E_{ads} + \Delta E_{ZPE} - T\Delta S \qquad (8)$$

where $\Delta E_{ZPE}$ represents the change of zero vibration energy, T and ΔS are the temperature and entropy change. Results are given in Figure 3d, the water solvation effect leads to about 0.03 eV downshift of ΔG (dashed lines). It is a wonder that full covered B/TM/B configurations possess a universal catalytic performance for HER. Especially, B/Cu/B, B/Pd/B and B/Al/B exhibit excellent catalytic performance with ultralow ΔG of -0.092 eV, 0.037 eV and 0.029 eV, respectively, which is comparable to Pt with ΔG of -0.075 eV.[2] B/Ni/B and B/Co/B exhibit larger ΔG of -0.274 eV and -0.773 eV, which are also acceptable for industrial electrocatalysis.

Further investigation on the HER performance of B/TM$_x$/B with different metal coverage x is conducted. As for B/Cu$_x$/B, B/Pd$_x$/B and B/Al$_x$/B, after eliminating deformed configurations, they exhibit excellent all-site defect-independent catalytic performance with extremely low ΔG ranging from -0.162 eV to 0.179 eV, -0.134 eV to 0.183 eV and -0.082 to 0.086 eV, respectively (Figure 3e). Detailed results are available in Figure S10~14 and Table S9~13 (Supporting Information). Besides, B/TM/B possess large surfaces, on which there are numerous adsorption sites for H, enabling efficient and parallel processes of HER catalysis. The readily available Cu, Al and B elements are abundant in the earth. Pd is half or two-thirds cheaper than Pt. Co and Ni are much more economical elements than noble metal as well. Consequently, B/TM/B are more economically viable and affordable than Pt.

In addition, corrosion of catalysts by electrolytes is an intractable problem in practical applications. In sandwiched B/TM/B, non-metallic borophene plays an ideal coating role with both chemical reaction inertness and God-given high conductivity. Without affecting the penetration of metal electronic states through borophene for catalysis, the intercalated metal can be effectively protected from the corrosion of electrolyte. These metal-borophene networks are expected to be ideal reaction platforms for electrocatalysis.

## 2.3. Electron distribution and orbital analysis

Charge distribution will change the electronic properties and chemical reactivity of B/TM/B, which crucially dictate its catalytic performance. Take charge density difference of B/Cu/B structure as en example (Figure 4a), in which electron accumulation and depletion are marked by yellow and blue, the intercalated metal atoms and boron triangles on the surface are the main sources of charge depletion. Due to the smaller electronegativity of metal atoms compared with B atoms, intercalated metal atoms tend to transfer electrons to borophene. As for boron triangles, metal atoms will induce a little charge depletion of adjacent B-B bonds. Electrons are mostly distributed between B/TM/B layers while part of electrons are captured by B-B bonds between boron hexagons, which activates their chemical activity (Figure 4b).

Charge density difference along Z axis is further obtained by projecting charge distribution after self-consistent calculation (Figure 4c), from which the charge transfer can be described quantitatively. Intercalated atoms are the major electron donors. The integral value of the curve indicates that 0.23 e has been transferred from Cu to B atoms. Boron triangles on top and bottom surfaces both lose 0.08 e. These electrons are distributed with a mirror symmetry with respect to metal atoms. Upon the Cu layer, nearly 3/4 charge (0.157 e) is located between B-Cu interlayer while the other 1/4 charge (0.046 e) is distributed around B-B bonds between boron hexagons, so does the other side. The number of charge transfer is determined by metal atoms (Figure 4d). The localization of these foreign electrons greatly activates B-B bonds and enhances their catalytic performance. Thus, H atoms, which prefer to be adsorbed at $B_1$ site, will participate in coupling to form a weak B-H-B bonding. Results for the other B/TM/B are given in Figure S15 (Supporting Information), which show similar outcomes.

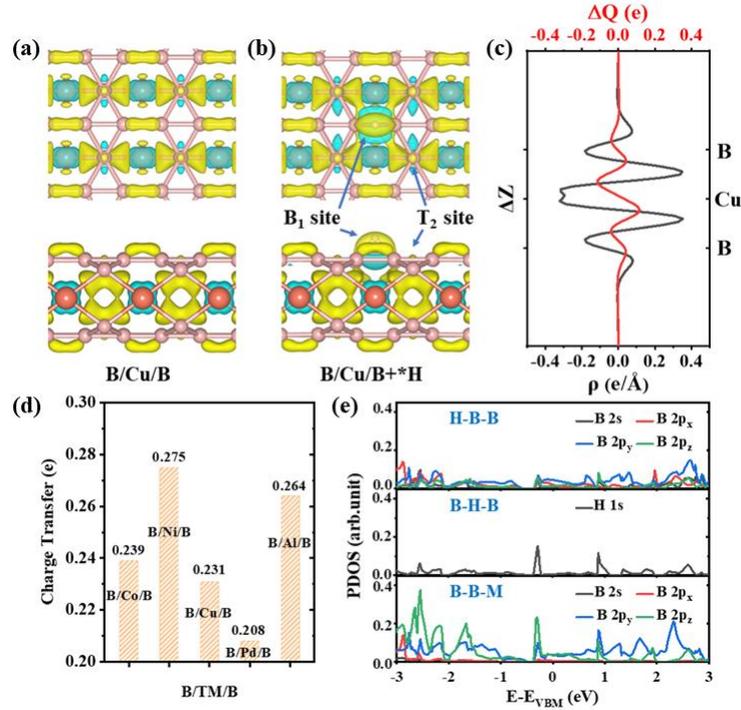

**Figure 4.** Top and lateral views of charge density difference of (a) B/Cu/B and (b) B/Cu/B+*H (isosurface level = 0.006 |e|/bohr$^3$). Yellow and blue indicate electron accumulation and depletion. (c) Charge density difference along Z axis of CuB$_6$. Integral values of the distribution curve (red lines) represent difference charge transfer. (d) Charge tranfer in B/M/B networks. (e) Partial charge density of states.

In order to further reveal the origin for excellent catalytic performance, partial charge density of states (PDOS, Figure 4e) is also investigated. As for B atoms in B-B bonds between boron hexagons, its ellipsoidally localized charge mainly comes from 2p$_y$ orbital, whose tendency is in strong accordance with 1s orbital of H atom. This is a decisive evidence to prove the favour of hydrogen adsorption. It is noteworthy that there is a little charge located at B atoms bonding with intercalated metal, which is contributed by 2p$_z$ orbitals of B atoms. This corresponds to T$_2$ site with a relative 1 eV higher adsorption energy, suggesting that it is not suitable for hydrogen adsorption according to the poor coupling with 1s orbital of H atom.

## 2.4. Thermal stability and kinetic behaviour

There is one crucial issue that whether sandwiched B/TM/B network could maintain considerable stability, so that it can be effectively applied. Therefore, ab initial molecular dynamics (AIMD) simulations are performed to provide more evidence of thermal stability. Time step is set to 1 fs during the 10 ps simulation. Taking B/Cu/B as an example, it maintains thermally stable under 1500 K but starts to distort once the temperature rises up to 1550 K (Figure 5a), indicating a decomposition temperature ($T_d$) of 1500 K. As shown in Figure 5b, among the five networks [B/TM/B (TM=Co, Ni, Cu, Pd) and B/Al/B], B/Al/B exihibits the highest $T_d$ of 2000 K, which is consistent with the previous reported 2080 K,[52] confirming the reliability of our calculations. Besides, B/Co/B and B/Ni/B present thermal stability under 1700 K and 1750 K, respectively. B/Pd/B exhibits a minimum $T_d$ of 1300 K, demonstrating the ability to maintain stable at high temperature as well. Thus, B/TM/B network is competent for strict conditions of industrial production. More details about the AIMD simulations are available in the methods part of the supporting information and Figure S16~20 (Supporting Information).

Migration of metal atoms was taken into consideration as well to reveal kinetic behaviour. Five possible pathways including vertical direction and in-plane cases are considered. As shown in Figure 5c, pathway 1 simulates the diffusion of single metal atom from the surface hollow site to interlayer intercalated site along Z axis. In pathway 2 and pathway 3, the single metal atom migrates along X axis and Y axis to the nearest interlayer equivalent sites, respectively. Pathway 4 and pathway 5 correspond to the surface conditions. Details are available in Figure

S21~25 and Table S14 (Supporting Information). Barrier along Z axis is much lower than the interlayer in-plane cases for all B/TM/B structures. Take surface migration into account, formation of B/Pd/B and B/Al/B may need introduction of temperature or pressure. Totally, pathway 1, where metal migration from surface towards interlayer of borophene bilayer, is more likely to occur.

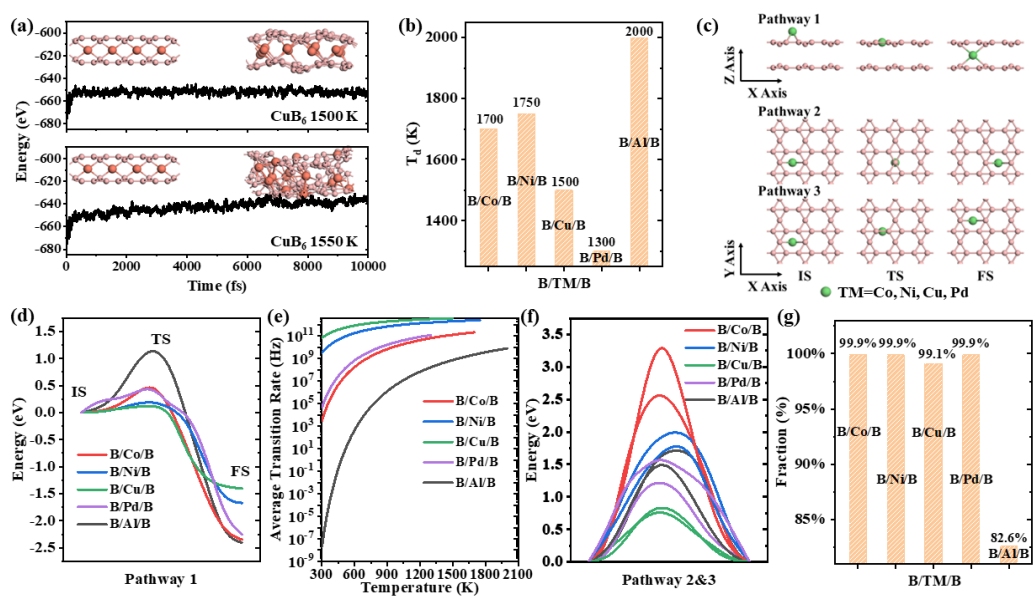

**Figure 5.** (a) AIMD of B/Cu/B under 1500 K and 1550 K. (b) Decomposition temperature ($T_d$) of B/TM/B. (c) Three pathways of metal migration in bilayer borophene. IS, TS and FS denote the initial state, transition state and final state, respectively. More details are available in Figure S21~25. (d) Barriers of migration and (e) Average transition rate along Z axis of B/TM/B. (f) Barriers of migration in XY plane. (g) The fraction of average transition rate along Z axis in total transition rate.

Barriers of metal migration along Z axis in B/TM/B are plotted in Figure 5d. Firstly, the energies of final states (FS) are much lower than that of initial states (IS) by 1.4 ~ 2.4 eV. This not only indicates that FS structure is much more stable than IS structure, but also reveals the irreversibility of migration along Z axis. Once intercalated, it is difficult for metal atoms to escape from borophene sheets. Metals in B/TM/B network migrate with different difficulty along the Z axis. The metal diffusion energy barrier is 0.13 eV in B/Cu/B, which is the lowest among the five networks. In B/Ni/B, an energy barrier of 0.21 eV exists. Co and Pd atoms need to overcome larger barrier, 0.57 eV and 0.50 eV, to migrate into interlayer. For comparison, Al atoms need to overcome a barrier as large as 1.24 eV to migrate.

The metal transition rate from surface to interlayer can be estimated by the formula $\nu^{*}\exp(-E/k_{B}T)$, where the prefactor $\nu$ is approximately equal to $10^{13}$ Hz, E denotes the energy barrier. $k_B$ and T are the Boltzmann constant and temperature, respectively. Average transition rates in B/TM/B along Z axis at tempreture from 300 K to their $T_d$ are given in Figure 5e. Temperature plays a decisive role in the migration of metal atoms. The transition rate rises significantly about several orders of magnitude with the increasing of temperature. Besides, the fraction of average transition rate along Z axis in total transition rate can be calculated. Both B/TM/B (TM=Co, Ni, Cu, Pd) exhibits ultrahigh proportion up to 99% while B/Al/B has lower proportion of 82.6% (Figure 5g), indicating the absolute dominance of metal migration along Z axis. In other words, the dissipation of intercalated metal at the edge of metal-borophene networks is suppressed. Our results demonstrate B/TM/B is highly thermal and kinetic stable.

3. Conclusion

The structure, metal coverage, formation, water splitting process, hydrogen adsorption, charge distribution, thermal stability and kinetic behaviour of B/TM/B (TM=Co, Ni, Cu, Pd) are systematically investigated. B/TM/B can work as ideal platform for water dissociation due to low water dissociation barrier. Especially, B/Cu$_x$/B, B/Pd$_x$/B and B/Al$_x$/B exhibit superior global catalytic performance (ΔG: -0.162 eV ~ 0.179 eV, -0.134 eV ~ 0.183 eV and -0.082 ~ 0.086 eV, respectively) comparable to Pt for HER. The wonderful catalytic performance is attributed to electron localization on the surface induced by charge transfer from intercalated metal to B atoms while intercalated metals are protected by wrapped borophene. In B/TM/B networks, the migration of metal atoms from surface to interlayer is absolutely dominant and irreversible, leading to the avoidance of metal dissipation at edge. Besides, B/TM/B is a high temperature resistant material with decomposition temperature of at least 1300 K. Thus, the 2D sandwiched B/TM/B structure is very robust while fulfils structural, energetic, thermal and kinetic stability. Overall, the advantages of B/TM/B such as high structural and thermal stability, low resistance to metal intercalation, effortless water splitting process, excellent defect-independent catalytic performance, cheapness and abundance of raw materials, make B/TM/B promising in electrocatalytic HER applications.

**Notes**

The Authors declare patent of invention, its application number is 202211702777.3.

**Corresponding authors:**

Junfeng Gao*: gaojf@dlut.edu.cn

**Author Contributions**

The manuscript was written through contributions of all authors. All authors have given approval to the final version of the manuscript. †Yuan Chang and †Hongsheng Liu contributed equally to the work.

## Acknowledgements

This work is supported by the National Natural Science Foundation of China (Grant No. 12074053, 91961204, 12004064), by the Fundamental Research Funds for the Central Universities (DUT21LAB112, DUT22ZD103, DUT22LK11). We also acknowledge Computers supporting from Shanghai Supercomputer Center, DUT supercomputing center, and Tianhe supercomputer of Tianjin center. Y-W Z acknowledges the support from Singapore NRF-CRP24-2020-0002 and Singapore A*STAR-SERC-CRF Award.